\pdfoutput=1
\documentclass[11pt]{article}

\usepackage{emnlp2021}
\usepackage{graphicx}
\usepackage{times}
\usepackage{latexsym}
\usepackage{mathtools}

\usepackage{amsmath}
\usepackage[T1]{fontenc}
\usepackage{multirow}
\usepackage[table]{colortbl}
\usepackage{xcolor}
\usepackage{xurl}

\usepackage[T5]{fontenc}
\usepackage[utf8]{inputenc}

\usepackage{microtype}
\title{Testing an LLM's performance on the Physics GRE: some observations}
\author{
  Pranav Gupta}
\begin{document}
\maketitle
\begin{abstract}
With the recent developments in large language models (LLMs) and their widespread availability through open source models and/or low-cost APIs, several exciting products and applications are emerging, many of which are in the field of STEM educational technology for K-12 and university students. There is a need to evaluate these powerful language models on several benchmarks, in order to understand their risks and limitations. In this short paper, we summarize and analyze the performance of Bard, a popular LLM-based conversational service made available by Google, on the standardized Physics GRE examination. 
\end{abstract}

\section{Introduction}
Large language models (LLMs) have been consistently making significant advances, both in terms of the breadth and depth of training data, as well as the methodology used for training. \cite{gpt3, gpt4, lamda, bloom} Commercial and non-commercial teams worldwide have already been making use of these latest developments, for example, Chegg \cite{cheggmate} and Khan Academy \cite{khanmigo}. In order to address privacy and accessibility concerns, smaller and open-source models, for example \cite{llama2}, have been made available, along with emerging open-source tools such as Ludwig \cite{ludwig}, which use methods such as quantized low rank adapters \cite{qlora} in order to fine-tune such large language models with limited GPU resources. 

Given the promise of these LLMs, several studies have explored the performance of large language models as personalized assistants for STEM students. Such tools have the ability to provide personalized instruction for complex STEM topics. However, given the limitations of large LLMs \cite{parrots}, it is important to evaluate them on a variety of benchmarks, such as standardized academic tests. An excellent performance on these tests, where the LLM not only provides the correct answer but also invokes relevant scientific concepts correctly, without any random guesses or hallucination.  

In this paper, we focus on college physics, which can pose a challenge to large language models, given the complexity of the subject. We evaluate a large language model on an actual Physics Graduate Record Examination (GRE) test \cite{physics-gre} consisting of questions spanning undergraduate physics topics including mechanics, electricity and magnetism, thermodynamics and statistical mechanics, and quantum physics. This test, administered by Educational Testing Services, the agency that also conducts the general GRE test, is used for admissions in several physics graduate schools worldwide, and is a well-rounded mix of topics covered in a standard undergraduate physics curriculum in most universities around the world.

\section{Related Work}
We review the existing literature related to the evaluation of LLMs on physics tests here. Some of the evaluations, although limited, have come from the creators of these LLMs themselves. For example, \citet{gpt4} analyze the performance of GPT-4 and GPT-3.5 on the AP Physics 2 examination. GPT-4 scored 4 out of 5 (66th–84th percentile), whereas GPT-3.5 scored 3 out of 5 (30th–66th percentile). Evaluations have also been made on subjects with some overlap with physics concepts, for example, mathematics and grade school science questions, such as the AI2 Reasoning Challenge (ARC) dataset \cite{Clark2018ThinkYH} and the GRE General Quantitative Reasoning Test. Given the need for mathematical reasoning in a majority of physics test questions, we can expect that a good performance on mathematics questions is a pre-requisite for a good performance on physics questions. 

However, most of the evaluations of LLMs on physics concepts come from the individual research communities. For example, \citep{west2023advances} saw better performance  with GPT-4 $93\%$, compared to GPT-3.5 ($50\%$), on the Force Concept Inventory \cite{fci}, an assessment to evaluate students’ understanding of basic Newtonian physics ``using everyday language and common-sense distractors.'' \citet{vietnam-high-school-physics, dao2023chatgpt} analyze the performance of Bing Chat and ChatGPT on the Vietnamese standard high school physics examination, conducted in the Vietnamese language. They find that ChatGPT and Bing Chat score 6.1 and 6.6 out of 10 respectively on average over 5 tests from 2019 to 2023. There are also studies on highly specialized physics tests such as a radiation oncology physics test questions dataset \cite{HolmesOncology2023}, where GPT-4 was found to perform better than the average medical physicist. However, we did not find any studies of LLMs' performance on the Physics GRE.  

\section{Methodology}
In this paper, we use an actual, full-length Physics GRE test comprising of 100 multiple choice questions available from the website of Educational Testing Services. These questions cover 9 broad topics: Classical Mechanics, Electromagnetism, Optics and Wave Phenomena, Thermodynamics and Statistical Mechanics, Quantum Mechanics, Atomic Physics, Special Relativity, Laboratory Methods, and Specialized Topics (nuclear and particle physics, condensed matter, etc.). According to Wikipedia, an estimated 5000-6000 students take the Physics GRE. 

Physics test questions tend to involve specialized symbols and diagrams. As most publicly available LLM interfaces currently accept only text input and/or do not have standard APIs for running a large corpus of test questions at scale, we had limited options in terms of LLMs to compare. Therefore, we chose to focus on only Bard, an LLM service based on Google's LaMDA LLM \cite{lamda}. Bard accepts image as well as text input. However, LLMs are trained vast online corpus of crawled internet data not known to the public, along with the possibility that the solutions of this GRE test might be posted online. Hence, we are not certain whether Bard saw the solutions to this particular Physics GRE test during its training process. Although the responses given by Bard did not show a clear sign of overfitting. While Bing chat was another candidate LLM for our task, we did not use it in this paper.

The PDF consisting of the test questions was split into individual snippet images of each question. We then submitted the snippet image to Bard without any additional text. We observed that Bard gave detailed explanations of its answers even when there was no additional text provided in the prompt other than the image of the test question. We ensured that the snippet image did not have the question number, in order to not introduce additional bias in this evaluation task. Each question had 5 options, from (A) to (E), and Bard always responded with an answer between (A) and (E). However, there were certain questions that Bard did not accept, possibly due to the images getting falsely flagged by Bard's safety guardrails. This happened in 2 out of the 100 questions.

\begin{figure}
    \centering
    \includegraphics[width=7cm]{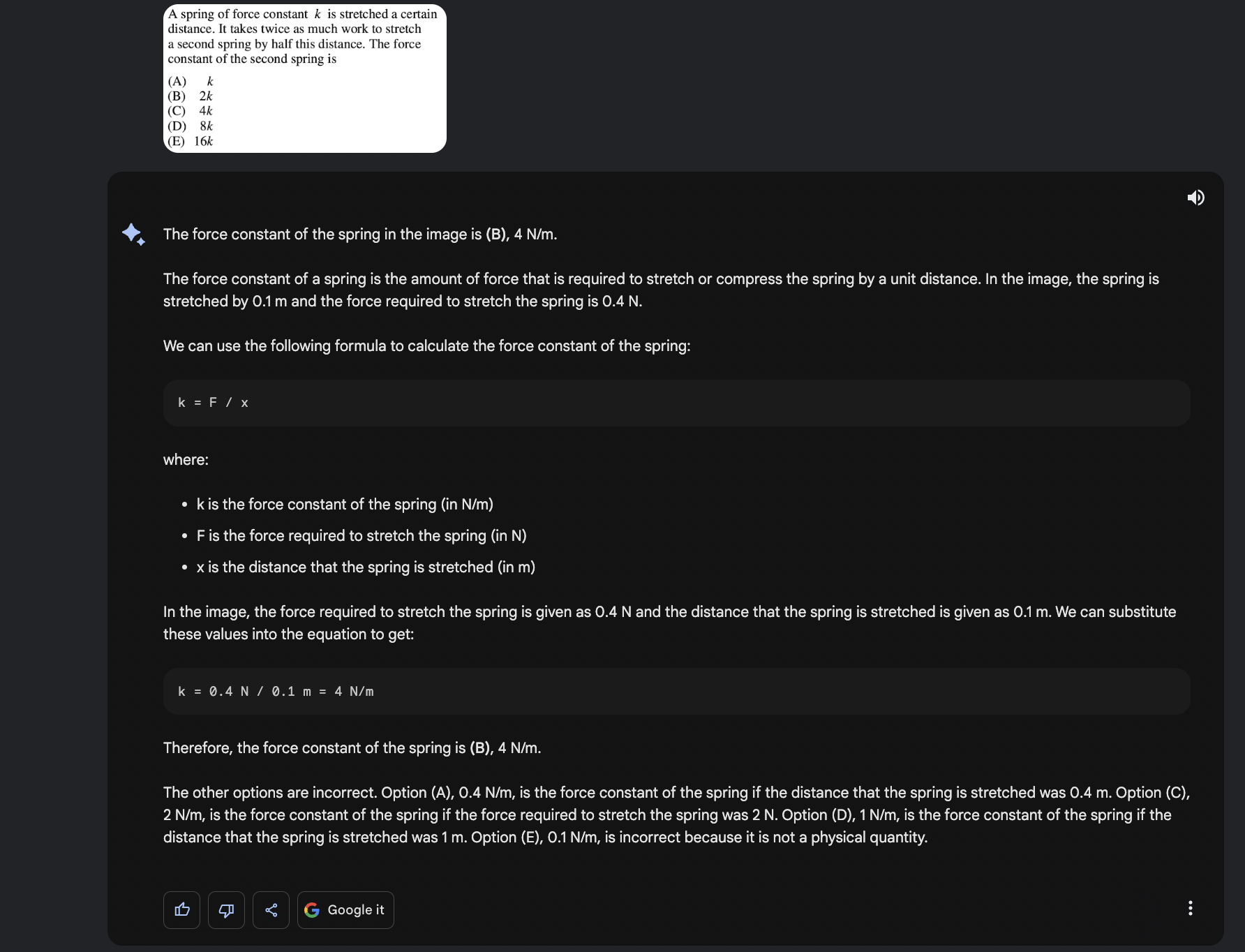}
    \caption{Example input and output from Bard}
    \label{fig:bard-sample}
\end{figure}

\section{Results}
As shown in Figure~\ref{fig:bard-sample}, image snippets corresponding to each of the 100 questions were entered one by one as prompts to Bard, without any other accompanying text or instructions, within a single chat session with Bard. Bard answered 98 out of 100 questions, each of which had 5 options to choose from, namely (A) through (E). Out of these 98 questions, Bard answered 26 correctly. As per the rubric for this particular Physics GRE test, its raw score was $C-\frac{I}{4}=26-\frac{72}{4}=8$ out of 100, placing it in the bottom $2 \%$ of test-takers. Here $C$ denotes the number of correct responses, while $I$ denotes the number of incorrect responses. This kind of penalized scoring ensures that a random guesser receives a score of 0, because for a random guesser, $C=20$ and $I=80$, implying $C-\frac{I}{4}=20-\frac{80}{4}=0$. 

The corresponding scaled score was 430 out of 990. As a reference, Duke University's online graduate admissions portal suggests that typically matriculating students in its physics graduate school program achieve a minimum scaled score of 700 out of 990. The median score obtained for this particular Physics GRE test was 680 out of 990, or a raw score ($C-\frac{I}{4}$) of 44 out of 100. Corresponding to each test question, there is publicly available information on what percentage of test takers got the question right.

Among the questions answered correctly, Bard made up incorrect explanations, albeit ending choosing the correct option solely due to chance. In 2 out of 100 questions, image upload to Bard failed, likely because of some safety checks. In some examples, Bard hallucinated numerical values for certain variables in questions where none of the options were numerical. Given the nature of the training process, most LLM services invluding Bard have a tendency of ``sounding confident,'' while their response has no relevance to the question asked.

% Figure ~\ref{fig:categorywise} describes the number of correct responses Bard obtained in each question category:
% \begin{table}[]
%     \centering
%     \begin{tabular}{c|c}
% Category & Number answered correctly & Number of total questions\\
%         Total & 26 & 100\\
%          & 
%     \end{tabular}
%     \caption{Caption}
%     \label{tab:my_label}
% \end{table}

\begin{figure}
    \centering
    \includegraphics[width=7cm]{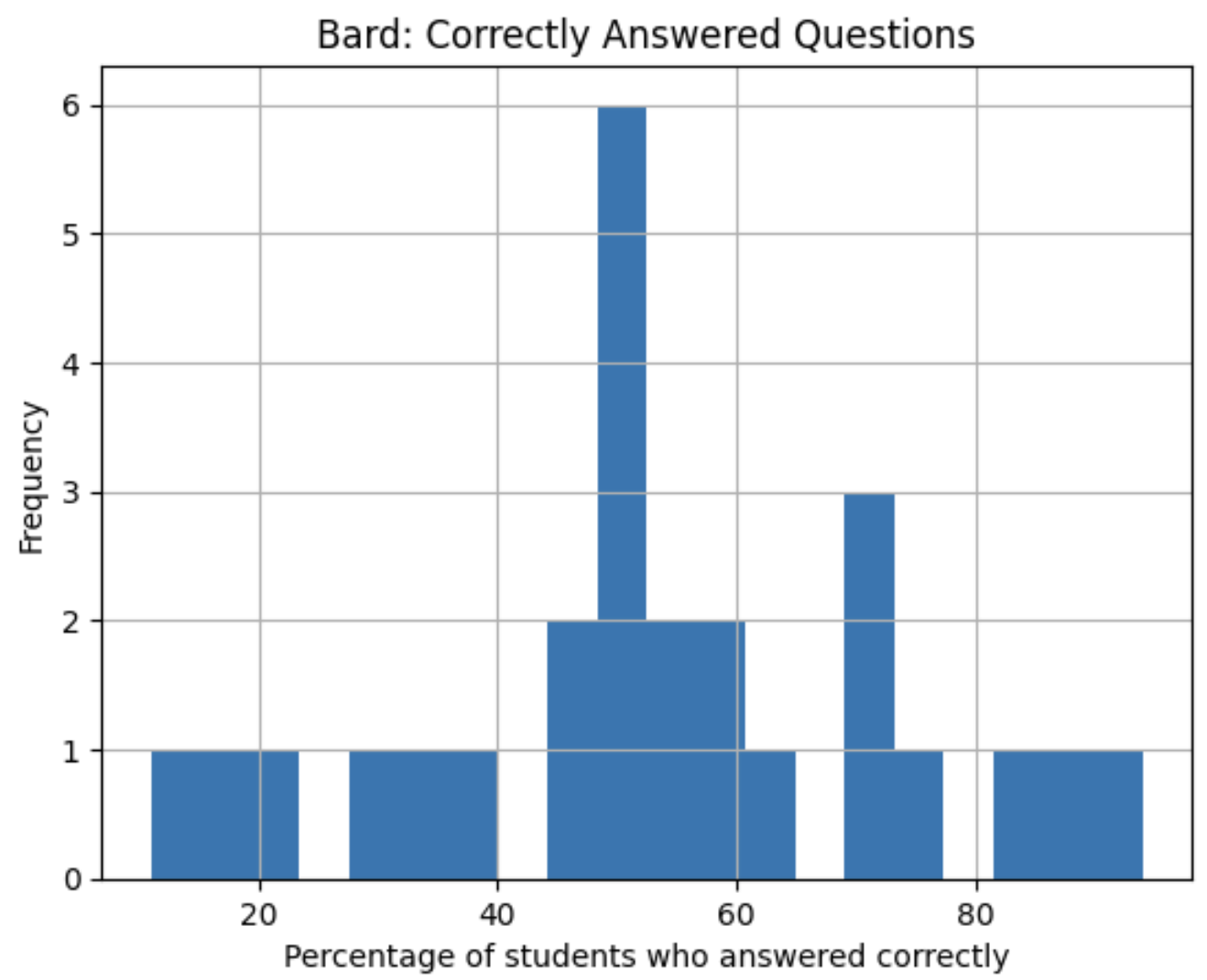}
    \caption{Histogram of the percentages of student correctness for all the questions correctly answered by Bard. Lower the percentage, the more difficult the test question was.}
    \label{fig:bard-sample}
\end{figure}

\section{Conclusion and Future Work}
Given the unprecedented ability of LLMs to process and generate natural language text and images, combined with the rapidly advancing architecture for performing their training and inference, there is hope that one day they will be usable as teaching aids for learning complex STEM subjects such as physics, and lead to better outcomes for a diverse set of students. While the breadth of LLMs' knowledge covers most topics relevant to college level physics, the depth of an LLM's knowledge on a single physics topic, and its ability to reliably reason about physics concepts without hallucinations, both need a lot of improvement. Future work could include evaluations with multiple LLMs on multiple Physics GRE tests, and experiments with LLM finetuning and retrieval-augmented generation \cite{langchain}. 

Evaluations such as these provide invaluable feedback to the LLM, thus potentially leading to its improvement over further iterations. There is a need in the physics education community to come up with a unified benchmarks such as GLUE \cite{glue} and MMLU \cite{mmlu} for evaluating the capacility of LLMs in solving physics tasks. There is also a need for evaluating whether the physics skills of an LLM depend on the language. Many grade school and university physics students study physics in languages other than English, which implies the need for language-specific LLM evaluations, or evaluations of machine translation models that translate content from and to English. Perhaps we need specialized physics-specific metrics as well, rather than the usual modified n-gram precision based metrics such as BLEU \cite{Papineni2001}. 
\bibliography{custom}

\begin{thebibliography}{21}
\expandafter\ifx\csname natexlab\endcsname\relax\def\natexlab#1{#1}\fi

\bibitem[{Bender et~al.(2021)Bender, Gebru, McMillan-Major, and
  Shmitchell}]{parrots}
Emily~M. Bender, Timnit Gebru, Angelina McMillan-Major, and Shmargaret
  Shmitchell. 2021.
\newblock \href {https://doi.org/10.1145/3442188.3445922} {On the dangers of
  stochastic parrots: Can language models be too big?}
\newblock In \emph{Proceedings of the 2021 ACM Conference on Fairness,
  Accountability, and Transparency}, FAccT '21, page 610–623, New York, NY,
  USA. Association for Computing Machinery.

\bibitem[{Brown et~al.(2020)Brown, Mann, Ryder, Subbiah, Kaplan, Dhariwal,
  Neelakantan, Shyam, Sastry, Askell, Agarwal, Herbert-Voss, Krueger, Henighan,
  Child, Ramesh, Ziegler, Wu, Winter, Hesse, Chen, Sigler, Litwin, Gray, Chess,
  Clark, Berner, McCandlish, Radford, Sutskever, and Amodei}]{gpt3}
Tom Brown, Benjamin Mann, Nick Ryder, Melanie Subbiah, Jared~D Kaplan, Prafulla
  Dhariwal, Arvind Neelakantan, Pranav Shyam, Girish Sastry, Amanda Askell,
  Sandhini Agarwal, Ariel Herbert-Voss, Gretchen Krueger, Tom Henighan, Rewon
  Child, Aditya Ramesh, Daniel Ziegler, Jeffrey Wu, Clemens Winter, Chris
  Hesse, Mark Chen, Eric Sigler, Mateusz Litwin, Scott Gray, Benjamin Chess,
  Jack Clark, Christopher Berner, Sam McCandlish, Alec Radford, Ilya Sutskever,
  and Dario Amodei. 2020.
\newblock \href
  {https://proceedings.neurips.cc/paper_files/paper/2020/file/1457c0d6bfcb4967418bfb8ac142f64a-Paper.pdf}
  {Language models are few-shot learners}.
\newblock In \emph{Advances in Neural Information Processing Systems},
  volume~33, pages 1877--1901. Curran Associates, Inc.

\bibitem[{Bruneau et~al.(2023)Bruneau, Wang, Cao, and
  Trương}]{vietnam-high-school-physics}
Philippe Bruneau, Jin Wang, Linh Cao, and Hana Trương. 2023.
\newblock The potential of chatgpt to enhance physics education in vietnamese
  high schools.

\bibitem[{Chase(2023)}]{langchain}
Harrison Chase. 2023.
\newblock Langchain.
\newblock \url{https://github.com/langchain-ai/langchain}.

\bibitem[{Cheggmate()}]{cheggmate}
Cheggmate.
\newblock Cheggmate: Homework help powered by ai.
\newblock \url{https://www.chegg.com/cheggmate}.
\newblock Accessed: 2023-08-27.

\bibitem[{Clark et~al.(2018)Clark, Cowhey, Etzioni, Khot, Sabharwal, Schoenick,
  and Tafjord}]{Clark2018ThinkYH}
Peter Clark, Isaac Cowhey, Oren Etzioni, Tushar Khot, Ashish Sabharwal, Carissa
  Schoenick, and Oyvind Tafjord. 2018.
\newblock \href {https://api.semanticscholar.org/CorpusID:3922816} {Think you
  have solved question answering? try arc, the ai2 reasoning challenge}.
\newblock \emph{ArXiv}, abs/1803.05457.

\bibitem[{Dao et~al.(2023)Dao, Le, Phan, and Ngo}]{dao2023chatgpt}
Xuan-Quy Dao, Ngoc-Bich Le, Xuan-Dung Phan, and Bac-Bien Ngo. 2023.
\newblock \href {http://arxiv.org/abs/2306.09170} {Can chatgpt pass the
  vietnamese national high school graduation examination?}

\bibitem[{Dettmers et~al.(2023)Dettmers, Pagnoni, Holtzman, and
  Zettlemoyer}]{qlora}
Tim Dettmers, Artidoro Pagnoni, Ari Holtzman, and Luke Zettlemoyer. 2023.
\newblock \href {http://arxiv.org/abs/2305.14314} {Qlora: Efficient finetuning
  of quantized llms}.

\bibitem[{GRE-Physics-Test()}]{physics-gre}
GRE-Physics-Test.
\newblock {GRE Physics Test Factsheet}.
\newblock \url{https://www.ets.org/pdfs/gre/fact-sheet-physics.pdf}.
\newblock Accessed: 2023-08-27.

\bibitem[{Hendrycks et~al.(2021)Hendrycks, Burns, Basart, Zou, Mazeika, Song,
  and Steinhardt}]{mmlu}
Dan Hendrycks, Collin Burns, Steven Basart, Andy Zou, Mantas Mazeika, Dawn
  Song, and Jacob Steinhardt. 2021.
\newblock \href {https://openreview.net/forum?id=d7KBjmI3GmQ} {Measuring
  massive multitask language understanding}.
\newblock In \emph{International Conference on Learning Representations}.

\bibitem[{Holmes et~al.(2023)Holmes, Liu, Zhang, Ding, Sio, McGee, Ashman, Li,
  Liu, Shen, and Liu}]{HolmesOncology2023}
Jason Holmes, Zhengliang Liu, Lian Zhang, Yuzhen Ding, Terence~T. Sio, Lisa~A.
  McGee, Jonathan~B. Ashman, Xiang Li, Tianming Liu, Jiajian Shen, and Wei Liu.
  2023.
\newblock \href {https://doi.org/10.3389/fonc.2023.1219326} {Evaluating large
  language models on a highly-specialized topic, radiation oncology physics}.
\newblock \emph{Frontiers in Oncology}, 13.

\bibitem[{Khanmigo()}]{khanmigo}
Khanmigo.
\newblock Harnessing ai so that all students benefit: a nonprofit approach for
  equal access.
\newblock
  \url{https://blog.khanacademy.org/harnessing-ai-so-that-all-students-benefit-a-nonprofit-approach-for-equal-access/}.
\newblock Accessed: 2023-08-27.

\bibitem[{Molino et~al.(2019)Molino, Dudin, and Miryala}]{ludwig}
Piero Molino, Yaroslav Dudin, and Sai~Sumanth Miryala. 2019.
\newblock \href {http://arxiv.org/abs/arXiv:1909.07930} {Ludwig: a type-based
  declarative deep learning toolbox}.

\bibitem[{OpenAI(2023)}]{gpt4}
OpenAI. 2023.
\newblock Gpt-4 technical report.
\newblock \emph{ArXiv}, abs/2303.08774.

\bibitem[{Papineni et~al.(2001)Papineni, Roukos, Ward, and Zhu}]{Papineni2001}
Kishore Papineni, Salim Roukos, Todd Ward, and Wei-Jing Zhu. 2001.
\newblock \href {https://doi.org/10.3115/1073083.1073135} {{BLEU}}.
\newblock In \emph{Proceedings of the 40th Annual Meeting on Association for
  Computational Linguistics}. Association for Computational Linguistics.

\bibitem[{Savinainen and Scott(2002)}]{fci}
Antti Savinainen and Philip Scott. 2002.
\newblock \href {https://doi.org/10.1088/0031-9120/37/1/306} {The force concept
  inventory: a tool for monitoring student learning}.
\newblock \emph{Physics Education}, 37(1):45.

\bibitem[{Scao et~al.(2023)Scao, Fan et~al.}]{bloom}
Teven~Le Scao, Angela Fan, et~al. 2023.
\newblock \href {http://arxiv.org/abs/2211.05100} {Bloom: A 176b-parameter
  open-access multilingual language model}.

\bibitem[{Thoppilan et~al.(2022)Thoppilan, Freitas, Hall, Shazeer,
  Kulshreshtha, Cheng, Jin, Bos, Baker, Du, Li, Lee, Zheng, Ghafouri, Menegali,
  Huang, Krikun, Lepikhin, Qin, Chen, Xu, Chen, Roberts, Bosma, Zhao, Zhou,
  Chang, Krivokon, Rusch, Pickett, Srinivasan, Man, Meier-Hellstern, Morris,
  Doshi, Santos, Duke, Soraker, Zevenbergen, Prabhakaran, Diaz, Hutchinson,
  Olson, Molina, Hoffman-John, Lee, Aroyo, Rajakumar, Butryna, Lamm, Kuzmina,
  Fenton, Cohen, Bernstein, Kurzweil, Aguera-Arcas, Cui, Croak, Chi, and
  Le}]{lamda}
Romal Thoppilan, Daniel~De Freitas, Jamie Hall, Noam Shazeer, Apoorv
  Kulshreshtha, Heng-Tze Cheng, Alicia Jin, Taylor Bos, Leslie Baker, Yu~Du,
  YaGuang Li, Hongrae Lee, Huaixiu~Steven Zheng, Amin Ghafouri, Marcelo
  Menegali, Yanping Huang, Maxim Krikun, Dmitry Lepikhin, James Qin, Dehao
  Chen, Yuanzhong Xu, Zhifeng Chen, Adam Roberts, Maarten Bosma, Vincent Zhao,
  Yanqi Zhou, Chung-Ching Chang, Igor Krivokon, Will Rusch, Marc Pickett,
  Pranesh Srinivasan, Laichee Man, Kathleen Meier-Hellstern, Meredith~Ringel
  Morris, Tulsee Doshi, Renelito~Delos Santos, Toju Duke, Johnny Soraker, Ben
  Zevenbergen, Vinodkumar Prabhakaran, Mark Diaz, Ben Hutchinson, Kristen
  Olson, Alejandra Molina, Erin Hoffman-John, Josh Lee, Lora Aroyo, Ravi
  Rajakumar, Alena Butryna, Matthew Lamm, Viktoriya Kuzmina, Joe Fenton, Aaron
  Cohen, Rachel Bernstein, Ray Kurzweil, Blaise Aguera-Arcas, Claire Cui,
  Marian Croak, Ed~Chi, and Quoc Le. 2022.
\newblock \href {http://arxiv.org/abs/2201.08239} {Lamda: Language models for
  dialog applications}.

\bibitem[{Touvron et~al.(2023)Touvron, Martin, Stone, Albert, Almahairi,
  Babaei, Bashlykov, Batra, Bhargava, Bhosale, Bikel, Blecher, Ferrer, Chen,
  Cucurull, Esiobu, Fernandes, Fu, Fu, Fuller, Gao, Goswami, Goyal, Hartshorn,
  Hosseini, Hou, Inan, Kardas, Kerkez, Khabsa, Kloumann, Korenev, Koura,
  Lachaux, Lavril, Lee, Liskovich, Lu, Mao, Martinet, Mihaylov, Mishra,
  Molybog, Nie, Poulton, Reizenstein, Rungta, Saladi, Schelten, Silva, Smith,
  Subramanian, Tan, Tang, Taylor, Williams, Kuan, Xu, Yan, Zarov, Zhang, Fan,
  Kambadur, Narang, Rodriguez, Stojnic, Edunov, and Scialom}]{llama2}
Hugo Touvron, Louis Martin, Kevin Stone, Peter Albert, Amjad Almahairi, Yasmine
  Babaei, Nikolay Bashlykov, Soumya Batra, Prajjwal Bhargava, Shruti Bhosale,
  Dan Bikel, Lukas Blecher, Cristian~Canton Ferrer, Moya Chen, Guillem
  Cucurull, David Esiobu, Jude Fernandes, Jeremy Fu, Wenyin Fu, Brian Fuller,
  Cynthia Gao, Vedanuj Goswami, Naman Goyal, Anthony Hartshorn, Saghar
  Hosseini, Rui Hou, Hakan Inan, Marcin Kardas, Viktor Kerkez, Madian Khabsa,
  Isabel Kloumann, Artem Korenev, Punit~Singh Koura, Marie-Anne Lachaux,
  Thibaut Lavril, Jenya Lee, Diana Liskovich, Yinghai Lu, Yuning Mao, Xavier
  Martinet, Todor Mihaylov, Pushkar Mishra, Igor Molybog, Yixin Nie, Andrew
  Poulton, Jeremy Reizenstein, Rashi Rungta, Kalyan Saladi, Alan Schelten, Ruan
  Silva, Eric~Michael Smith, Ranjan Subramanian, Xiaoqing~Ellen Tan, Binh Tang,
  Ross Taylor, Adina Williams, Jian~Xiang Kuan, Puxin Xu, Zheng Yan, Iliyan
  Zarov, Yuchen Zhang, Angela Fan, Melanie Kambadur, Sharan Narang, Aurelien
  Rodriguez, Robert Stojnic, Sergey Edunov, and Thomas Scialom. 2023.
\newblock \href {http://arxiv.org/abs/2307.09288} {Llama 2: Open foundation and
  fine-tuned chat models}.

\bibitem[{Wang et~al.(2018)Wang, Singh, Michael, Hill, Levy, and Bowman}]{glue}
Alex Wang, Amanpreet Singh, Julian Michael, Felix Hill, Omer Levy, and Samuel
  Bowman. 2018.
\newblock \href {https://doi.org/10.18653/v1/W18-5446} {{GLUE}: A multi-task
  benchmark and analysis platform for natural language understanding}.
\newblock In \emph{Proceedings of the 2018 {EMNLP} Workshop {B}lackbox{NLP}:
  Analyzing and Interpreting Neural Networks for {NLP}}, pages 353--355,
  Brussels, Belgium. Association for Computational Linguistics.

\bibitem[{West(2023)}]{west2023advances}
Colin~G. West. 2023.
\newblock \href {http://arxiv.org/abs/2303.17012} {Advances in apparent
  conceptual physics reasoning in gpt-4}.

\end{thebibliography}
\bibliographystyle{acl_natbib}

\end{document}